# Citizen-informed scenarios to shape energy demand reduction strategies

Andy Yuille[1,2], Oliver Broad[3], John Barrett[4], Christian Brand[5], Alice Garvey[4], Jonathan Norman[4], James Price[3], Steve Pye[3], Rebecca Willis[1,2]

[1]Lancaster University, Bailrigg, Lancaster, United Kingdom.

[2]University of Manchester, Oxford Road, Manchester, United Kingdom.

[3]University College London, Gower Street, London, United Kingdom.

[4]University of Leeds, Woodhouse Lane, Leeds, United Kingdom.

[5]University of Oxford, Wellington Square, Oxford, United Kingdom.

## Graphical abstract

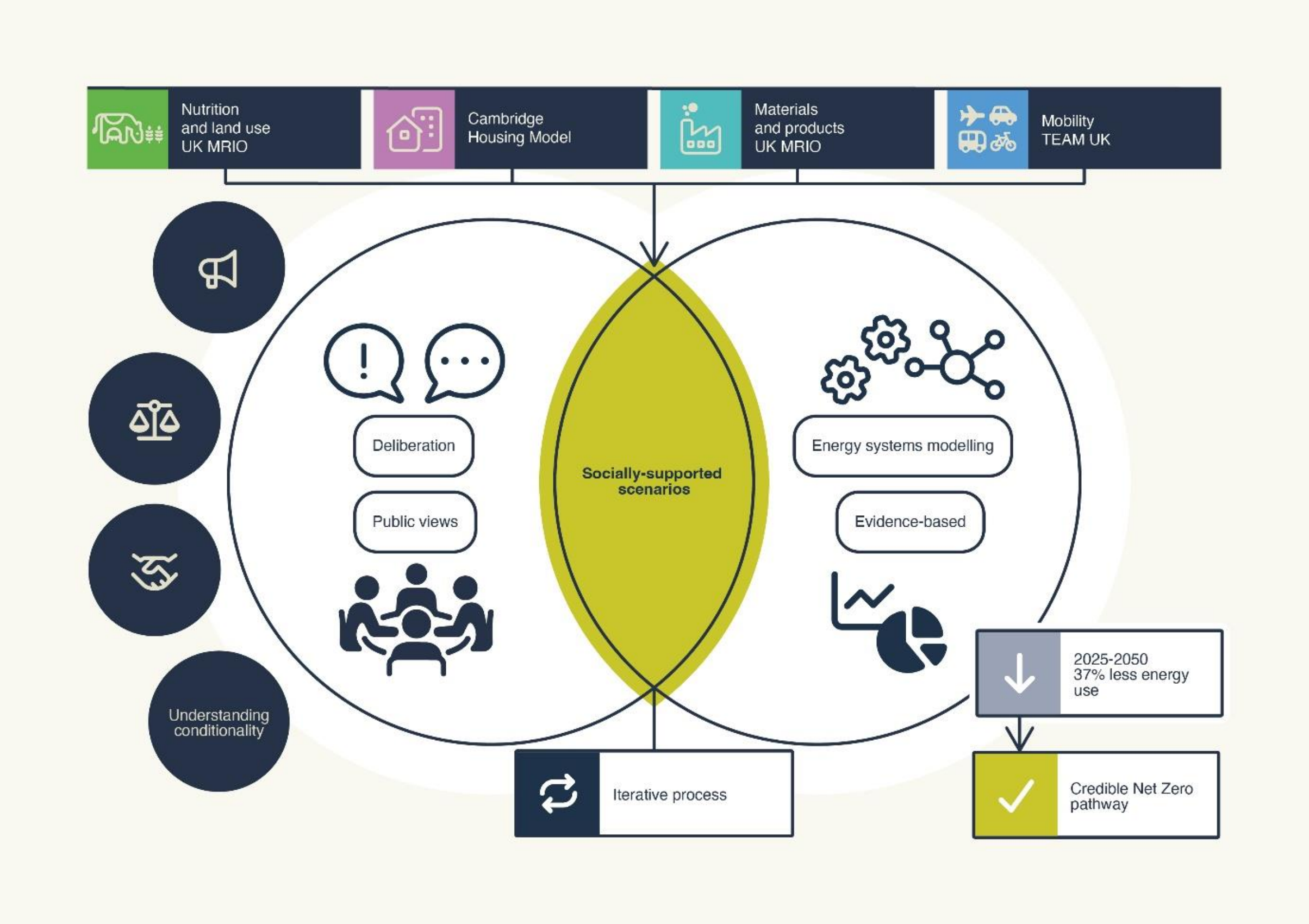

## Abstract

Growing evidence has demonstrated the important role that energy demand reduction can play in achieving net zero. Despite this, the adoption of such measures in government strategies often remains limited, in part based on perceived public resistance to change. Using a novel combined deliberative-modelling approach, which fills a key gap in the literature, we find this is unfounded. Citizens both recognise the importance of and support demand-side interventions to achieve the UK's net-zero goals. This support, however, is conditional. If policy interventions can make demand-side choices practical and fair, we find citizens are prepared to reduce energy demand by at least 37% between 2025 and 2050, providing 20% of the abatement required to meet net-zero targets. Policymakers could thus unlock a wide range of demand reduction measures by acting on a more comprehensive understanding of public views and needs. Moreover, our approach offers a powerful framework, applicable in other contexts, to underpin policymaking with socially-informed quantitative modelling.

## 1 Introduction

Demand-side measures could reduce global and European energy demands by 40% and 50% respectively by 2050 compared to 2020 [1,2], significantly contributing to net-zero strategies [3,4]. Corresponding analysis for the UK estimated energy demand reduction (EDR) potential of over 50% by 2050 compared to 2020 [5], a scale mirrored in other UK net-zero pathway studies that emphasize the link between EDR and social change [6].

Notwithstanding, governmental focus on EDR strategies that go beyond efficiency remains rare [7,8], despite growing evidence that demand-side measures support other policy objectives, such as energy security, health and wellbeing [9,10]. While policymakers recognise the importance of exploring demand-side focused scenarios [11], this has not been effectively translated into policy.

There is limited research on the reasons for the asymmetry between supply- and demand-side energy policies [12,13], but one prominent driver appears to be policymaker concern about resistance from a public unwilling to accept change to their everyday lives [14,15]. This reflects a well-documented 'perception gap', whereby politicians and the public both underestimate levels of public and political support for climate action [16-18]. However, despite consistently high concern about climate change and broad support for climate action, support for specific policies is often contested and, in some cases, declining [19].

One approach to bridging this gap relies on integrating participatory and analytical methodologies. While there is a growing literature on participatory energy systems modelling [20,21], it remains in its infancy and has not so far dynamically engaged the public using representative citizens' panels. Indeed, McGookin et al. note that "*the systematic participation of citizens is a key gap in energy systems modelling*" [20], and that particularly in national-scale studies, this results in "*failing to provide real-world 'on the ground' knowledge, experiences, perceptions and values*". Modelling is central to informing public policies [22], but is typically not directly informed by public engagement, or shaped by public views [23].

Citizens' panels are increasingly used to incorporate public views into developing policy advice, but are rarely integrated with modelling [24]. In its recent recommendations on setting the UK's Seventh Carbon Budget, the UK government's statutory advisers gathered evidence from a citizens' panel to complement traditional modelling. But even such positive examples fall short of full methodological integration [25].

This research advances this field by providing such a methodological framework. Innovating on existing work, the approach integrates a citizens' panel, whose members are selected through civic lottery or 'sortition' to ensure they are broadly reflective of the wider population [26], with energy systems modelling. This representative citizens' group iteratively informs scenario development across the full modelling process, from the bottom-up sectoral picture through to national-scale whole-system modelling. It is the first attempt to robustly quantify socially acceptable, technically achievable levels and types of energy demand reduction in a national pathway to net zero, and to provide detailed insights into the conditionality necessary to achieve those reductions. As such, it begins to address the pressing need for an improved understanding of the conditions which could enable public support for EDR, and the extent to which it could be implemented.

The result of the process was a panel-led scenario that reduced UK energy demand by 37% between 2025 and 2050, with 84% of panel members supporting at least this level of reduction. Stronger efforts on EDR and expanding clean electrification were preferred to heavier reliance on carbon dioxide removals to reach net zero. A variety of measures to reduce energy demand and emissions were supported across a range of sectors. However, critically, this support was conditional on governments enacting adequate enabling policies. These included financial support, but also regulatory, infrastructural, information-based, and industrial and community collaboration actions.

Our approach demonstrates the feasibility of combining deliberative research with energy systems modelling analysis to generate quantified scenarios with a demonstrable social mandate. Climate policymaking requires scenario modelling to set targets and guide policy. This will increasingly involve action to reduce emissions that entails changes to people's everyday lives [27]. Designing effective policies will thus increasingly require a more detailed and nuanced understanding of public views and values. Such an understanding will also help extend the range of policy options perceived as viable.

## 2 Using deliberative research to inform models of energy futures

Energy system models (ESMs) are crucial tools for energy and climate policymaking at national and global scales, but struggle to dynamically represent energy demand. Because it is 'societally-induced', energy demand is not intrinsic to an energy system but is shaped by wider societal, economic and political factors. It is actually demand for outcomes, like travel or heating, which in turn are shaped by physical infrastructures, social norms and practices [28-30].

In parallel, intentionally or not, government actions also directly and indirectly influence energy demand. Royston, et al. [31] use the term 'invisible energy policy' to describe how policy in areas as diverse as city planning, public service provision, fiscal policy and infrastructure investment can influence demand for energy. Societally-induced demand therefore differs by location and changes over time, often slowly, sometimes quickly. The twentieth century emergence of a car-based transport system from the interplay of technological change, public investment in road networks, changing settlement patterns, land-use planning, and evolving social expectations, exemplifies this process [32,33].

These demand dynamics present two distinct problems for ESMs. First, on complexity, ESMs, as bounded systems, struggle to capture potential future social and economic changes that influence will demand [5]. Second, on legitimacy, irrespective of their limitations, ESMs fail to represent whether proposed EDR interventions to achieve energy system outcomes are realistic or desirable.

Poor representation of societally-induced demand in ESMs can create a vicious circle. With modelling and policy support misrepresenting or excluding opportunities to change demand, policymakers do not have good evidence on the likely implications of these changes [34]. The absence of this evidence could lead to missed opportunities or unexpected outcomes, e.g. increasing future energy needs, especially if policymakers' assumptions excessively constrain the exploration space [35]. Conversely, if ESMs incorporate better representations of societally-induced demand, policymakers are presented with both a wider suite of options for purposive change, and a better understanding of indirect energy implications of policy decisions.

To better represent societally-induced demand, and the options for wider change, ESMs need to be informed by social research, to assess whether possible changes are plausible and legitimate. There is a range of existing social research that models can draw on for this evidence, including polling, survey data, and policy evaluation from other countries. However, to develop a reliable picture of the potential for change, it is important to establish what people would support or accept under different conditions and in different scenarios.

Therefore, building on Barrett, et al. [5], we use deliberative research to introduce insights on the plausibility and legitimacy of EDR measures into the modelling framework. As members of a citizens' panel, participants are offered information about demand-side options; reflect on, question, and deliberate these; and offer their views on which approaches they would support, under what conditions, and why [36]. Modelled results at both sectoral and whole-system levels are presented back to the panel for evaluation, discussion, feedback and refinement. The approach thus also reshapes modelling practice through iteration between modelling and deliberation. Citizens work alongside technical experts to co-produce scenarios throughout their development, making those scenarios socially legitimate as well as technically robust.

## 3 Deliberation on sectoral ambition

The citizens' panel included 40 members, selected through sortition [26] to be broadly reflective of the UK public across age, gender, ethnicity, disability, occupation, political views, climate concern, and residence in rural or urban areas. In eight blocks of meetings over two years, panel members considered possibilities for energy demand and emissions reduction, in terms of 'how we heat our homes', 'how we get around', 'what we buy', and 'what we eat'. Future scenarios for each sector were iteratively modelled based on panel discussions and voting. Sectoral scenarios were then integrated into a whole energy system model and a further process of deliberation and refinement generated a panel-led scenario. Figure 1 summarises the process which is fully described in Methods.

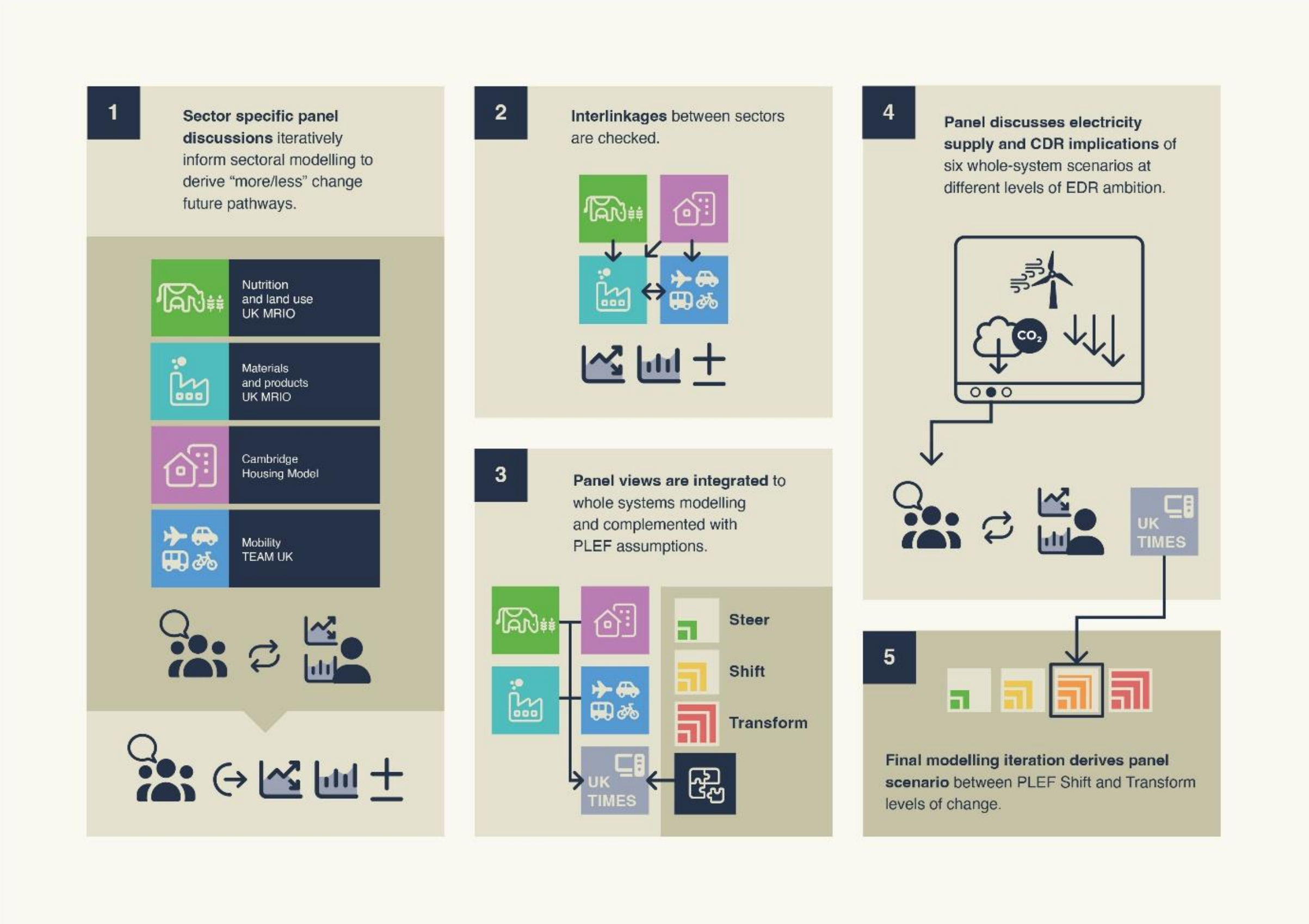

**Figure 1: Deliberative-modelling process overview**

Members showed a strong willingness both to change their everyday practices to reduce energy demand and emissions, and to back government action to support those changes. Figure 2 illustrates the levels of EDR supported by the panel in the sectors they considered, relative to three of the Barrett et al. Positive Low Energy Futures scenarios (hereafter "PLEF scenarios") [5]. For buildings and transport, the largest energy consuming sectors, the level of EDR is similar to that observed in the Shift scenario. In the PLEF scenarios, Shift represents an ambitious set of interventions across the whole economy, and is considered the most ambitious low demand scenario except for Transform. For industry, an even greater level of ambition is determined by the panel, exceeding the PLEF Transform scenario. For the food sector, including agriculture and food processing, energy reductions were much lower than PLEF despite energy efficiency gains, largely due to different panel choices around UK-based production driven by an objective of greater self-sufficiency.

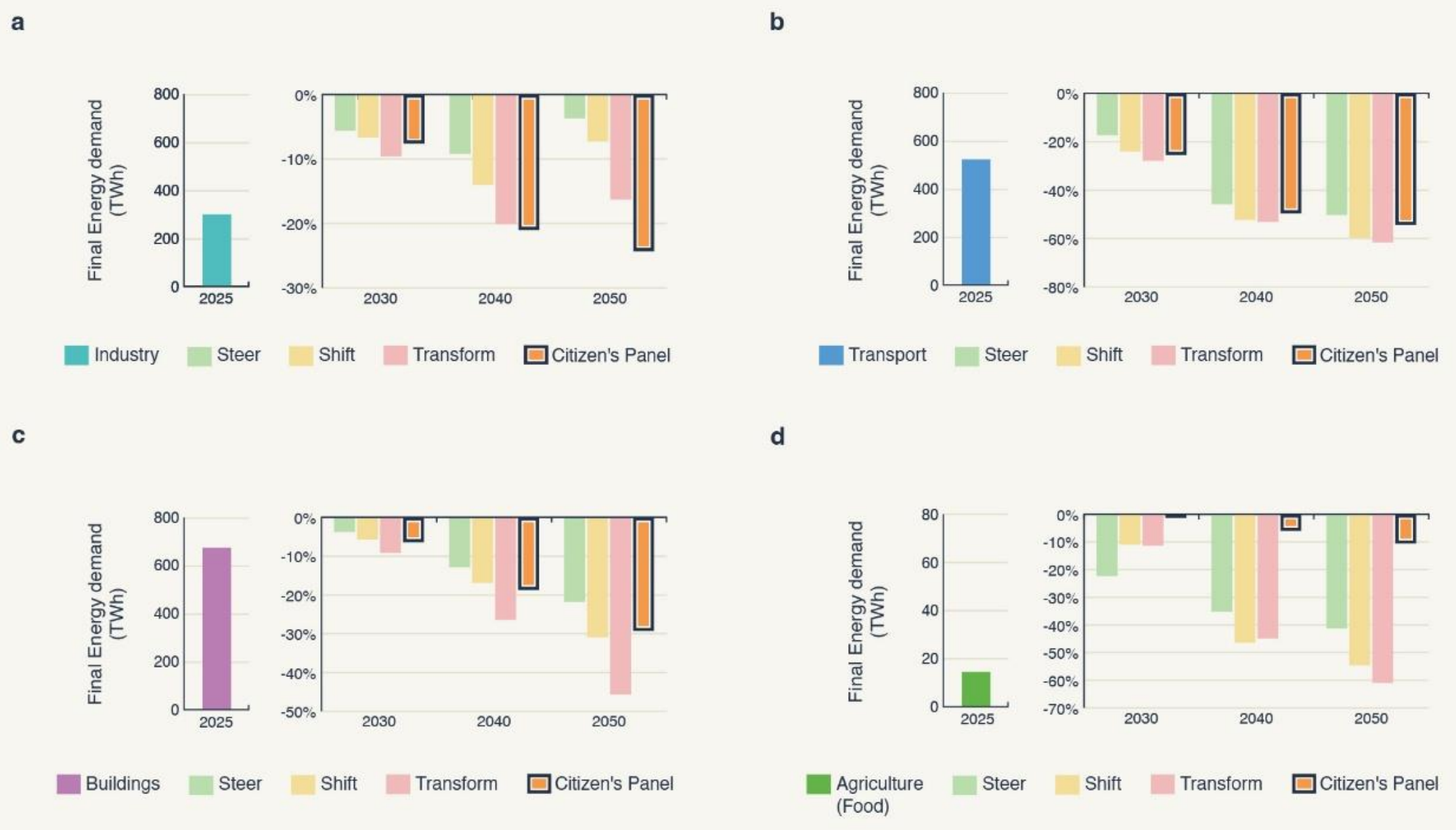

**Figure 2. Energy demand reduction by sector in the Citizen Panel scenario compared to Barrett et al. (2022) scenarios for selected years relative to 2025.** Sector panels include **a)** Industry, **b)** Transport, **c)** Buildings, **d)** Food. Note that the demand reduction for PLEF scenarios (Steer, Shift and Transform) reported in Barrett et al. (2022) was relative to a 2020 base year, but has been adjusted in this figure to be relative to 2025. All scenario results reported here, both PLEF and Citizen's Panel, are sourced from the whole system model, UK TIMES, but informed based on inputs from range of sectoral modelling (as described in Methods). Food, a relatively small energy consuming sector, shows smaller reductions relative to PLEF scenarios due to a range of factors, including a stronger focus on emission reductions than energy, and alternative choices not considered in PLEF that have higher energy intensity, including greater local production.

This confirms recent studies suggesting there has been a missed opportunity in developing policies targeted at reducing energy demand [5,7]. It goes beyond previous research to show the extent and type of EDR that could gain public support, and crucially the conditionality underpinning that support (summarised in Table 1).

Members supported EDR measures, and government action to enable them, across all sectors. Support for extent and rate of change and enabling interventions varied between sectors and between actions within sectors, discussed further in Section 5. Although complete consensus was rare, members were able to agree on conditions that would overcome objections and barriers for a large majority in relation to most actions.

| | **Actions** | **Conditions** |
|---|---|---|
| **How we** | • Widespread shift to heat pumps and other low-carbon heating | • Financial support for upfront costs<br>• Lower electricity prices |
| **heat our homes** | • Phase out new gas boilers by 2035<br>• Insulate existing homes, prioritising minimal disruption<br>• Build new homes to higher standards<br>• Increase flexibility in home energy use, while protecting people with inflexible energy use | • Trusted, accessible, and regulated information, advice, and engineers<br>• Stronger consumer protections<br>• Regulation to accelerate transition to new homes being net zero<br>• More access to flexible tariffs, with ultimate end-user control over use |
| **How we get around** | • Reduced total vehicle kilometres by:<br>- Shifting some journeys from private car to public and shared transport and active travel<br>- Reducing distance of some journeys<br>- Reducing frequency of some journeys<br>• Reverse the trend towards larger cars<br>• Continue shift from ICE vehicles to EVs<br>• Limit growth in flying, without unfairly disadvantaging people on lower incomes | • Transport solutions tailored to the needs of places and communities<br>• Better integrated, more affordable and reliable public and shared transport<br>• Safe, attractive walking and cycling routes<br>• Flexibility in when and where people work<br>• Locate and design development to reduce car dependence<br>• Car-sharing and on-demand transport schemes encouraged, led by community need and control<br>• Limit car size, and incentivise and encourage driving smaller cars<br>• Improved and expanded EV charging capacity<br>• Limit airport expansion and business flying, and tax those who fly most |
| **What we buy** | • Buying less but better, saving money, space and waste<br>• Products designed to last, and to be reused, repaired and remade<br>• Share and hire infrequently used items<br>• Develop community initiatives such as repair cafés and car clubs | • Standards requiring products to be repairable, upgradeable, and durable<br>• Sharing economy products and services accredited as safe and reliable<br>• Sharing and circular economy services as convenient, or cheaper, than buying new<br>• Build trust, respect and community spirit<br>• Leverage institutions like schools, clubs, and workplaces |
| **What we eat** | • Reduce the total amount of red meat and dairy eaten in the UK<br>• Develop local food communities<br>• Dedicate less land to livestock farming<br>• Lower emissions and more nature-friendly livestock farming<br>• Significantly less food waste | • Practical education, information, and engagement across society<br>• Individual diets continue to vary widely and freely<br>• Eating habits change slowly over time<br>• Healthier, low-emissions food in schools, the public sector and workplaces<br>• Regulate food industry on advertising, labelling, promotions and discounts<br>• Support low-income families<br>• Reform farm subsidies and import standards Strong policy action to reduce food waste at all levels |

**Table 1: Key actions supported by panel members, and conditionality of support.** Actions and conditions listed here garnered support from the majority of panel members.

## 4 Balancing EDR with energy supply and CDR

Achieving net-zero emissions across the whole economy requires some combination of EDR, decarbonising energy supply, and carbon dioxide removal (CDR). While the latter is increasingly expected to be necessary to compensate for residual emissions [27], the relative balance of these three 'pillars' is contested, with significant implications for pathway design, investment, and infrastructure buildout. Following deliberation on sectoral ambition, the panel next considered the role of EDR and the relative balance of these mitigation pillars in achieving net zero at the system level.

To introduce these pillars and the trade-offs between them into the deliberative process, six representative whole-system pathways were produced using the UK TIMES whole energy system model. Two levels of ambition for EDR measures were produced through iterative panel discussions and sector modelling. These were combined with the three PLEF scenarios [5] to describe increasing levels of ambition for those EDR measures not discussed by the panel.

A results dashboard facilitated panel interaction with these scenarios, supporting members to understand the implications of different levels of EDR for power sector size and CDR, and assess their preferred balance based on their understanding of the costs, risks, opportunities and benefits associated with each [37]. The final balance from the resulting panel-led scenario (CP) is presented in Figure 3a.

Energy demand under CP could be reduced by 37% between 2025 and 2050 (Figure 3b). While 16% of panel members thought this was too prominent a role for EDR, the remaining 84% conditionally supported at least this level, with 26% preferring that EDR go even further. This level of demand reduction would provide a fifth of total emissions reduction required for the period 2025-50. While CP remains below the most ambitious PLEF scenario, *Transform*, it slightly exceeds the proactive *Shift* scenario, showing high levels of member support for EDR as part of the UK net-zero and energy strategies. This is broadly comparable to the UK Climate Change Committee's 'Balanced Pathway' scenario, the modelled decarbonisation trajectory which underpins the recommendations of the UK Government's statutory advisors. This highlights that the scale of energy demand reduction in statutory advice to the UK Government on achieving net zero is aligned with that proposed by the panel [6].

Deep whole system electrification should accompany ambitious EDR and could provide 62% of emissions reductions required to reach net zero. Under CP, the level of generation almost doubles, rising from 280 TWh in 2025 to 545 TWh by 2050 (Figure 3c). Panel members recognised the interplay between EDR and system build-out, noting that while lower demand implies a smaller power system, and thus lower costs, deeper electrification could mitigate the need for what some perceived as excessive levels of change to everyday lives. Others raised concerns about the achievability and cost of rapid roll-out of low carbon energy infrastructure, and risks such as competition between renewables and food production.

Members accepted the need for some CDR to mitigate residual emissions. Under CP, CDR contributes 72 MtCO2 of removals, or 18% of the cumulative mitigation effort required to meet net zero in 2050. The high cost and significant uncertainty around scaling new technologies led to panel members preferring to shift the balance of CDR's expected contributions strongly in favour of land-based systems over engineered solutions. In 2050, CP shows 41 $MtCO_2$ (57%) of emissions removals provided through soil carbon and forestry (Figure 3e). Engineered removals total 31 $MtCO_2$ (43%) in 2050 (Figure 3d). While members supported further research into removals options, they strongly believed that engineered solutions that are unproven at scale should not be relied upon in policy or modelling to meet targets.

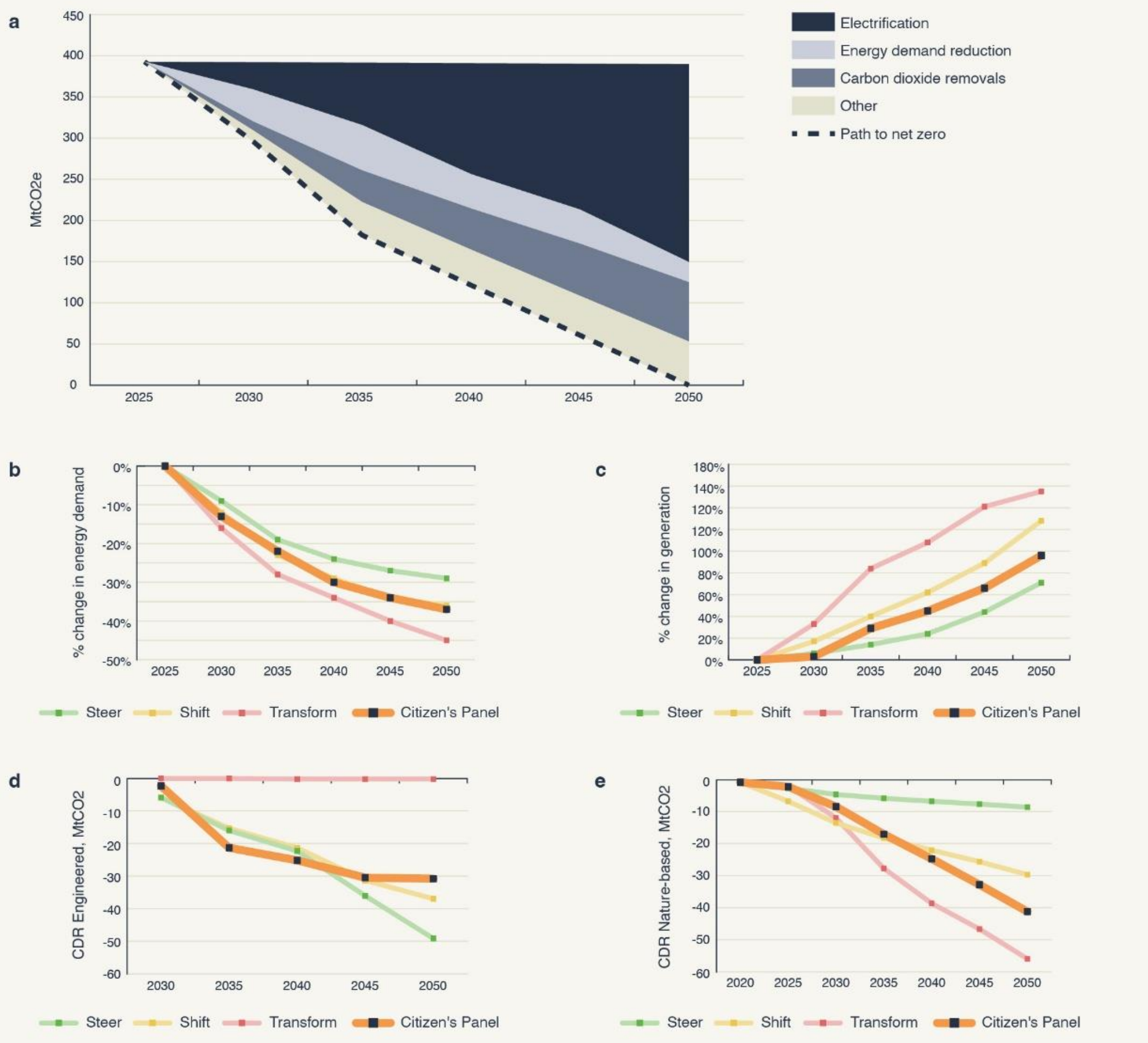

**Figure 3. Contribution of key types of mitigation measures to UK's net-zero target. a)** Decomposition of the Citizen's Panel (CP) scenario using the LMDI approach, showing the contribution of mitigation measures to meeting net-zero emissions by 2050, 2025-50. Contributions to mitigation are relative to the 2025 base year emissions level. Category definitions are as follows: *Electrification* includes decarbonisation of end use energy, primarily from the electricity system but also via other low carbon fuels. The category also includes reduction in emissions from electricity generation; *EDR* includes energy efficiency gains and reductions in energy service demand levels e.g. lower transport or heating demand; *CDR* includes engineered and land-based removals; *Other* includes emissions not captured in the above categories, including emissions from agricultural non-CO2 GHGs and upstream extractive sectors. Note that some measures, such as a heat pump, contribute to both *electrification* and *EDR*. Emissions reductions for electrification reflect the lower carbon intensity of electricity versus the fuel replaced e.g. gas, while EDR benefits are associated with a lower amount of energy being used due to efficiency gains. **b-e** show how the CP scenario compares to the Steer, Shift, and Transform scenarios (Barrett et al. 2022 5) across the three mitigation 'pillars' including **b)** energy demand reduction, **c)** electricity generation level, and **d)** engineered & **e)** natured-based carbon dioxide removal.

## 5 Discussion

This research provides evidence for two important findings: firstly, that citizens support a wide range of EDR measures as a means of achieving climate targets and can provide in-depth insights into the conditions required to underpin them; and secondly, that integrating deliberative research into energy systems modelling can deliver quantified EDR scenarios with a clear social mandate.

### 5.1 Conditional support for EDR

Previous research has shown that EDR can contribute significantly to emissions reduction [5] and that policymakers see the need for it [11]. This research completes the picture, confirming that citizens would support such a strategy. Panel members supported a 37% reduction in energy demand between 2025 and 2050. This is comparable in scale both to the 45% previously shown to be technically feasible between 2025-2050 [5], and to the advice provided by the UK Climate Change Committee via their 'Balanced Pathway' scenario [6]. demonstrating clear support for EDR to form a key element of UK net-zero strategy.

However, support for EDR is conditional on how policy is developed, designed and communicated. Members displayed nuance, sophistication and willingness to engage with contrasting viewpoints in their questioning and interpretation of evidence, recognition of barriers to change, and proposed solutions.

Yet tensions do remain. As representative groups of citizens, deliberative "mini-publics" rarely come to complete consensus, and this is to be expected when discussing complex and controversial policy issues. Identifying key areas of and reasons for disagreement is a core outcome of deliberation, recognising the legitimacy, credibility and nature of disputed values, beliefs and choices [38]. This dynamic was reflected in panel discussions where vocal opposition to measures from a minority of members with deeply-held beliefs sometimes dominated conversation, but were not reflected in voting outcomes. For example, while 11% of members were firmly opposed to heat pumps, a majority were willing to get one, with 64% supportive of and 25% willing to "live with" a phase-out of new gas boilers by 2035, under the right conditions. Outcome legitimacy rests more on fairness and inclusion than on consensus [39].

Policymakers should therefore not avoid contested areas, but rather approach them with care. Decisions need not be supported by everyone but must show deep consideration for the breadth of public sensitivities, ensuring policies are designed and communicated to meaningfully address motivations and concerns to maximise conditional support [19].

Despite these tensions, members supported EDR measures across all sectors, with variations across intervention type, scale, and implementation rate both within and between sectors. For example, there was majority support for higher taxes on larger cars and frequent flights, but strong opposition to taxing meat. Changing diets was considered a gradual process that should be driven by 'softer' measures, and unlikely to reach the *Transform* PLEF scenario level [5]. In contrast, some changes to what we buy (such as loaning infrequently used items and increasing the ratio of repair over replacement) could be achieved more quickly, exceeding the ambition of the *Transform* scenario. In all cases, outcome fairness - understood broadly to encompass differential impacts between e.g. income groups, dis/ability, location, and households and the private sector - and the protection of vulnerable groups was a key driver behind support or opposition for particular changes and policies.

Freedom of choice was highly prized, with a desire to see policy consistently making it easier, more attractive, and 'more normal' for people to make low-energy and low-emissions choices. Members recognised that EDR requires a combination of upstream (industry) and downstream (public) action. Where new regulation was required, they largely favoured focusing this upstream, e.g. regulating the food industry on labelling, advertising, and promotions, and the electronics industry on lifespan, repairability and modularity. They favoured changing public behaviour largely through society-wide information and education, incentives, and structural changes to shift norms and expectations. Examples include 'normalising' lower-meat diets through school and public sector catering and supermarket marketing and promotions; improving the integration, reliability and affordability of public transport; and lowering electricity prices.

Support for change was conditional on consistent and sustained government leadership and accountability. Members recognised the vital role of government in creating policies, setting targets, raising and spending money, and communicating with the public. However, they had little confidence in governments making the right decisions or implementing them well. They believed that governments' ability to take an appropriately long-term view was constrained, and were sceptical about policies surviving changes in administration. They did not believe that trade-offs and social impacts were adequately considered or communicated, or that the public had sufficient input into policymaking. Integrating deliberative research into policymaking is one route to rebuilding this trust and confidence.

**5.2 Implications for modelling and policymaking**

This research demonstrates that including citizen deliberation as an input to energy systems modelling and scenario development is both feasible and beneficial.

Previous research has shown that citizens' panels selected using sortition can grapple with complex issues and contribute meaningfully to policy development [39]. In this case, we have developed a methodology in which citizen deliberation not only provides meaningful input into scenario development at sectoral and whole system level, but also into the evaluation and refinement of scenarios, delivering the kind of transparent, collaborative, iterative approach that the literature has identified as missing from participatory energy system modelling [21, 40].

In doing so, they have provided vital insights into the conditions, trade-offs and contingencies involved in policymaking.

Close working between social researchers and the modelling team enabled outcomes of panel discussions to be iteratively incorporated into model runs. However, this required considerable interdisciplinary understanding and patience, given the very different skillsets and methodologies involved.

The research provides a vital demonstration of the social mandate for EDR policies and a quantified description of the levels of energy and emissions reductions that could realistically be achieved, underpinned by detailed insights into conditions of support. This provides richer and more nuanced guidance for policymaking than other means such as polling or surveys. It can give policymakers greater confidence to implement EDR measures, as well as signposting which may require careful handling.

This approach, if widely used and communicated to citizens, can help to build trust in government, countering scepticism about net-zero strategies for, and beyond, energy demand reduction [41,42]. It strengthens the democratic process through providing more sophisticated understandings of citizen views than voting patterns, surveys, or focus groups, and provides a necessary foil to economic interests whose voices are over-represented in policymaking [43-45].

A comprehensive, nuanced understanding of the views and experiences of citizens will be vital in developing future climate strategy. Combining deliberative research with quantitative modelling can produce scenarios which are both technically feasible and supported by citizens, and can extend the range of policy options perceived as viable. As such, these approaches should be widely incorporated into policymaking processes.

# Methods – Online

## Introduction

This work builds on previous multi-step methods for assessing national low energy demand futures. Barrett et al. (2022) [5] establishes the method on a case study of the UK, describing an academic view of “technically feasible” Energy Demand Reduction (EDR) given different levels of societal, political, and infrastructural transformation. Sharmina et al. (2025) [11] expands the approach, using a policy-led co-creation process to assess drivers of future change, thus moving from theoretical EDR potential to politically salient future demands. Here, we take the third step, pivoting from policy to citizen perspectives, expanding the method to assess the social mandate for EDR. To this end, we have developed a novel “deliberative-modelling” approach whereby a representative Citizens’ Panel co-designs the scenario analysis of possible future changes to energy demand levels in the UK.

In this section, we first describe the deliberative citizens’ panel framework on which the public engagement element of this work is based. We then detail the modified stepwise approach inherited from Barrett et al. (2022) [5], replacing researcher-led narrative building and sector modelling with a citizen-based, iterative, deliberative framework. To the best of our knowledge, this represents the first attempt to iteratively combine the deliberations of a representative citizens’ panel with energy systems modelling, marking a significant advance for both fields. Finally, we include a short description of the sectoral and whole systems models used, and discuss key limitations of our work to date.

## The deliberative process

There is an increasing awareness of the need for more dialogic and inclusive engagement between citizens and states on complex and controversial issues such as climate and energy policy. One approach that has recently gained traction is the “deliberative mini-public" (DMP), also known as citizens’ assemblies or panels. Following a review of 282 DMPs in 18 member countries, the OECD advocates institutionalising them to improve decision-making and strengthen democracy [24,46]. Since 2018 there have been over 200 climate-related DMPs in Europe, from local to national level [47]. This approach is valued for generating meaningful insights into public views on specific issues, given appropriate evidence and the time to consider it with people from different backgrounds [39].

Substantively, this can support more robust policymaking by engaging deeply with diverse and often unheard views and motivations; procedurally, it can strengthen the integrity of policymaking processes; ethically, it adheres to the principle of people shaping policies that will affect them; and instrumentally, it is considered to increase public trust, boosting social acceptance and perceived legitimacy for policymaking [48].

The deliberative process has four key stages:

1. Selection – the panel is selected to reflect the wider population it represents. This UK case study used the 'gold standard' civic lottery process known as 'sortition' [26] to select 40 members. Selection factors included age, gender, occupation type, ethnicity, disability, rural or urban residence, concern about climate change, and political views. Additionally, all members lived within 90 minutes of central Manchester, to facilitate in-person meetings.
2. Learning – members are presented with balanced, high-quality information about the topic, which they reflect on, discuss, and question. As in this case, guidance on balanced evidence is often provided through an independent advisory group.
3. Deliberation – trained facilitators guide members through structured discussions and activities to exchange, reflect on, and develop their views based on their interactions with experts, commentators, and each other.
4. Producing conclusions – these may reflect consensus (collective support), consent (active support from some and acceptance by others), and/or division (identifying and clarifying the spectrum of views). Records of deliberation capture the reasoning underpinning conclusions and give insights into motivating values, judgements, experiences, thought processes, emotions, and preferences.

Here we take the highly novel step of combining the DMP format with energy systems modelling. Outputs from panel deliberations iteratively informed scenario development as described in the following sections, to derive quantified, technically feasible pathways to net zero with a demonstrable social mandate.

Panel meetings were held over two years across 8 thematic "Deliberative Modelling Blocks" (DMBs). Each DMB consisted of two evening online and one weekend in-person workshop where deliberative stages 2-4 were undertaken. DMBs 2-5 included sector-specific modelling, and a fourth online workshop at which results from the quantitative analysis and deliberative outputs that underpinned it were presented back to members for review. This resulted in the panel meeting 36 times in 28 2-hour workshops online and 8 full-day in–person workshops in Manchester. (Further DMB information available in SI).

Balanced information on the topics was provided to members in plenary by a range of expert stakeholders. Discussions and activities were led in groups of up to 8 by a trained facilitator. Questions were asked of speakers both within small groups as they arose, and in plenary following group question generation and prioritisation. All DMBs captured both collectively and individually articulated views, using methods such as individual voting and collective rating and prioritising, backed up with reasoned justifications and dissenting opinions.

Data was captured in a variety of ways to enable triangulation of findings. Key discussion points were written on post-its, pre-designed flipcharts (e.g. using matrices, mindmaps, and scales), and their electronic equivalents. Votes were taken using 'dot voting' and Mentimeter. Rating and scoring used Likert scales. Photos were taken of all in-person material outputs, and online equivalents downloaded. These were then transcribed into Word and Excel documents to facilitate coding and analysis. Group conversations were recorded and transcribed under confidentiality agreement. Qualitative data was analysed using thematic and narrative analyses and used to interpret quantitative data.

The following sections describe the 5-step method used to combine the DMP with integrated sector and whole systems energy modelling. These steps mirror those presented in the process diagram Figure 1. Step 1 focused on key sectors of the economy in relation to energy demand and GHG emissions, using the connected themes of "what we eat", "what we buy", "where we live" and "how we get around". Each theme was the focus of one DMB. Step 2 considered sector inter-dependencies. Step 3 combined sectoral model outputs into the UK TIMES whole system model, generating 6 new scenarios of varying levels of EDR ambition. In Step 4, the panel considered whole-system outcomes and trade-offs from previous sectoral discussions, informing their final Citizens' Panel-led scenario (CP). In Step 5, a final DMB reviewed CP and synthesised high-level conditionality for its successful implementation.

## Step 1 – Sectoral Panel Blocks for deliberative-modelling

Sectoral DMBs share an overall structure, consisting of (1.1) knowledge exchange and deliberation, (1.2) analysis of panel views and modelling, and (1.3) pathway result feedback and deliberation. While not described here, sectoral DMBs were preceded by a set of introductory panel meetings which familiarised members with climate science and impacts; carbon targets, budgets and key emitters; modelling and scenarios; existing EDR research; and the panel process and focus.

(1.1) Knowledge exchange and deliberation took place within one week, following the deliberative process stages 2-4 described in the previous section. Each DMB theme included discussion of up to four demand reduction topics. These aligned with widely-discussed sectoral EDR and emissions reduction solutions; identified knowledge gaps; and/or significant energy demand or emissions impacts. DMBs were tightly scoped to allow rich deliberation on each topic. For example, in "what we buy", topics included car clubs and journey-sharing schemes; longer life and more repairable electronics; conscious clothing consumption; and hiring or borrowing infrequently–used products. (Full topic list presented in Table SI1).

Information on each topic was introduced by both internal project members and external experts. The mix of speakers accounted for expertise, recognised authority, or key roles held within the theme, and included both academic experts and professional practitioners. In “what we eat”, for example, the panel heard from a farmer, a carbon accounting consultancy, and representatives from a major UK supermarket, a circular economy NGO, and a sustainability think-tank. They also heard from academic experts on the personal, social and cultural meanings of food; trends in UK diets and their health, cost, social and emissions implications; and consumer-, retail- and producer-facing policy options to support dietary change.

After knowledge introductions, facilitated group discussions generated questions for speakers, and deliberative exercises helped members connect discussion topics with their lived experience. These had three objectives. First, to help name and describe motivations for, and barriers to, supporting EDR opportunities within the topic. Second, to articulate reasoned conditions which would partially or fully overcome barriers. Third, to assess the extent to which enabling conditions could address concerns and increase support. For example, implementing transport conditions significantly increased overall support for shifting some journeys from car to other modes and/or closer to home; implementing heat pump conditions significantly increased speed of uptake, while only modestly increasing overall support; and implementing car club conditions had relatively little impact due to lack of confidence that conditions could, in practice, be met, and to deeper-seated attachments to car ownership. Two outcomes, “with” and “without” conditions, describe futures representing levels of uptake for each topic that translate into futures with “more” and “less” change in energy demand.

(1.2) Analysis of panel views and modelling started with producing a discursive summary of members’ views regarding the DMB sector. This provided the context within which the quantified analysis was understood and interpreted. Levels of support for each change in demand were quantified through voting, providing “more” and “less” change outcomes, which were then used to inform sectoral modelling. For example, the maximum roll-out of heat pumps into existing UK homes was indexed on the proportion of panel members willing to adopt the technology. The speed at which this roll-out takes place was based on whether support was expressed for immediate, near future, or much later adoption. If members’ stated barriers to uptake were mitigated, both the share of supporters, and the timeframe for action, shift. These two different outcomes describe the trajectories used to build the lower and higher modelled demand scenarios of building energy use.

This process was repeated for each of the topics discussed in each DMB. Thus each DMB produced two quantified sectoral scenarios of future energy demand that translate panel members’ values, judgements and preferences at “more” and “less” levels of change.

Because the structure and characteristics of each sector shape how energy demand and emissions are generated, each one was analysed using a dedicated, sector-specific, simulation model. These are described briefly in the “models used” section below, and mirror the tools used in Barrett et al. (2022) [5] and Sharmina et al. (2025) [11].

(1.3) Finally, pathway result feedback and deliberation in each DMB occurred through an online workshop. A summary report of key qualitative findings was sent to members in advance. Short qualitative and quantitative presentations were prepared. The first described discussion outcomes, including key conditions identified as necessary to facilitate more change in EDR, and the concerns they were intended to address. The second showed the quantified implications for future UK energy demand and emissions of higher and lower change scenarios.

Members collectively reflected, questioned speakers, and assessed whether the scenarios accurately reflected their discussions, conclusions, and expectations. They were then guided through conversations exploring particular findings in more detail to enhance understanding of conditionality or help refine modelled scenarios. For example, in “what we eat”, the panel considered under what conditions additional regulations on food companies would be acceptable; in “what we buy” they discussed the factors underlying the strong resistance to car-sharing schemes. Following this meeting the modelled scenarios were refined where necessary, and a final DMB summary report was sent to panel members.

## Step 2 – Sectoral interdependencies

The four themes discussed with the panel covered “what we eat”, “what we buy”, “how we heat our homes” and “how we get around”. These correspond to the four end use sectors of nutrition, materials and products, shelter, and mobility.

Understanding and mapping interconnections between sectors ensures that the final narrative and corresponding modelling analysis are internally consistent [5,11]. The objective is to ensure that relevant changes in one sector are reflected in assumptions included in another. For example, the requirement for new vehicle production (including vehicle size) arising from the transport modelling was aligned with UK demand for such vehicles in the materials and products modelling.

By design, the scope of each DMB, in terms of EDR topics, was kept tight to fit within time and deliverability constraints (see Step 1). Where model assumptions or outcomes in one sector were discussed, corresponding links to other sectors were explicitly considered and shaped based on panel views. These links are shown in Figure 1. Where topics or sectoral characteristics were excluded from the DMBs, consistency across sectors was based on sourcing input from Barrett et al. (2022) [5] scenarios (see Step 3).

## Step 3 – Combined Panel and PLEF whole-systems framework

Here we integrated the panel-informed sector analysis into the whole energy system modelling framework. This process mirrors Step 4 described in Barrett et al. (2022) [5] with key adjustments accommodating citizens' panel input.

First, the views expressed by the panel regarding the sectors they have discussed were integrated into the UK TIMES whole energy system model. Specifically, outputs from sectoral modelling from each DMB describing "more" or "less" change were combined to provide an overall picture of "more" and "less" change in energy demand. Using established connections between sectoral models and UK TIMES described in Barrett et al. (2022) [5] and Sharmina et al. (2025) [11], these levels of change were translated into "more" and "less" change scenario variants for the whole system. For example, UK TIMES emissions pathways describing future agricultural practice were aligned with outputs from the nutrition modelling, and constraints on heat pump deployment in the residential sector were adjusted to reflect different levels of panel ambition.

Second, areas not discussed by members were complemented using PLEF scenario levels of change. DMBs 2-5 focused on specific sectors, and, within a sector, on specific topics (changes) in relation to EDR. Discussing additional sectors, or further topics, from previous work [5,11] was not practical. Yet undiscussed sectors or topics still stand to have significant impact in shifting energy demand across the whole system. To allow panel members to review these measures at an aggregate level we applied Steer, Shift and Transform PLEF scenario assumptions from Barrett et al. (2022) [5] alongside panel stated views. By identifying all the options included in PLEF but not discussed in DMBs, we created three "Background" futures against which the panels' "low" and "high" ambition for change might play out.

When combined, these two assumption sets generated six whole energy system EDR scenarios, each combining one level of Background change with one level of panel ambition (e.g. Steer+Low, Steer+High, Shift+Low, etc.).

## Step 4 – Panel-led trade-offs for system decarbonisation

These six scenarios were presented via a bespoke interactive dashboard, available at https://ourenergyfutures.org/, that was introduced to members through a combination of presentations and hands-on exploration. Guided and self-guided activities, in-person, in groups of three used laptops provided by the research team to build an understanding of scenario results.

The purpose of the dashboard was to help members understand the interactions between three "pillars" of net zero: varying levels of EDR, and their implications for power generation, and for land-based and engineered carbon dioxide removals (CDR). Mirroring other DMBs (Step 1), members were presented with information about land-based and engineered CDR, the scale and type of clean power generation and infrastructure that may be required, how these levels relate to varying levels of EDR, and how the three pillars interconnect to underpin different pathways to net zero. Following structured discussions analysing the costs, benefits, risks and opportunities posed by power generation and different types of CDR independently, members then considered the key issues associated with these two pillars in relation to each other, and their possible trade-offs with EDR. This enabled members to come to a view on the overall balance of these three core "pillars" of net zero that they would support.

The deliberation in this DMB led to adjustments in modelling assumptions, used to derive a final citizens' panel-led scenario (CP). CP is thus based on both detailed consideration of EDR and emissions reduction opportunities in specific sectors, and the acceptability of different aggregate levels of EDR, CDR, and power generation at the whole system level.

## Step 5 – An integrated CP scenario for EDR in the UK

The scenario that most closely reflected members' views ("more change" panel-led sectoral scenarios, combined with background assumptions from the PLEF Shift scenario) was modified to respond to the outcomes from Step 4 to derive CP. This involved rebalancing CDR in the scenario away from engineered to land-based removals, accounting for system level impacts where appropriate. Ultimately, CP sees 57% of CDR delivered by land-based systems.

In the final DMB, CP was presented to the panel both qualitatively, with narrative reference to its effects on everyday life, and quantitively, in terms of its effects on UK energy use and emissions. Particular focus was given to the most recent analysis adjustments, i.e. the challenges and implications of a large increase in land-based CDR. Again, reflecting Step 1, members reflected on and discussed this information and questioned the speakers.

Facilitated discussion then assessed the extent to which CP represented an appropriate outcome of their discussions over the course of the full panel process, raising remaining member concerns, and their views on how they might choose to deviate from it, e.g. by putting more or less emphasis on particular 'pillars' or elements within pillars. Following this discussion, members voted individually on how well CP reflected panel discussions; how well it aligned with their personal views now in terms of EDR; the balance between the three 'pillars'; and the balance between land-based and engineered CDR.

Finally, panel members developed and agreed on a set of ten high-level principles for achieving EDR fairly and effectively, derived from the detailed conditions identified as necessary to enable action in individual sector DMBs and thus necessary to underpin implementation of CP.

## Models used

The modelling approach used in this study builds on work presented in Barrett et al. (2022) [5] and Sharmina et al. (2025) [11]. Sectoral simulation models were used to analyse nutrition, materials and products, shelter, and mobility, which are reflective of the four thematic DMBs. Outcomes from these models were integrated into the UK TIMES whole energy system optimisation framework.

For nutrition, a hybrid physical input-output model draws on observed economic data of intermediate and final demand from the UK Multi-Regional Input-Output (MRIO) model, as well as a detailed construction of average UK diets based on established nutrition surveys [49,50]. For materials and products, a hybrid model combined supply chain representations from the UK MRIO model, with a bespoke model of the UK construction sector. Resource efficiency opportunities involving actions by UK final consumers and industries were modelled as changes in final or intermediate (industry) demand. The impact of such changes on UK industrial output were captured through supply chain modelling. For shelter, the Cambridge Housing Model (CHM) is a micro-simulation model of existing UK homes that couples a Standard Assessment Procedure (SAP)-based home energy performance calculator with English Housing Survey data [51]. Over 16,000 representative dwellings with descriptions of building type, envelope characteristics, heating system, climatic zone, and others, are weighted and run against SAP building physics calculations to derive final energy consumption and $CO_2$ emissions. For mobility, the Transport Energy Air pollution Model for the UK (TEAM-UK) is built around the UK National Travel survey data [52], and simulates passenger demand based on travel indicators. It presents a transport-energy-environment system-modelling framework.

Finally, outputs from sector models were integrated into the UK TIMES whole energy system model of the UK. Used both in academic and policy setting roles, UK TIMES is a linear cost-optimisation model that offers an internally consistent and technology-explicit environment within which to develop future decarbonisation pathways of the UK. It includes five end use sectors (agriculture, residential, services, transport, industry), and derives required energy flows and infrastructure needs to meet their end use demand considering resource extraction, primary and secondary fuel production (including the power sector). Here, sectoral constraints and thresholds were used to mirror sector model pathways, with demand trajectories aligned to follow outputs from the sectoral models. UK TIMES includes all carbon budget, NDC, and Net Zero targets for the UK, and accounts for competition between sectors for resources.

This integration ensured that long-term system-wide implications of individual sector changes, as well as the possible energy and emissions trade-offs between these sectors, were accounted for in a fully internally consistent analysis framework.

These modelling tools are described in detail in their respective literatures [53-60], and their use as a combined framework for studying EDR is presented in Barrett et al. (2022) [5] and Sharmina et al. (2025) [11].

## Limitations

The framework presented here combines for the first time deliberative citizen science with integrated energy systems modelling. It is interdisciplinary by design, and, over two years, brought together research teams across multiple universities, third sector deliberative panel experts, expert and specialist speakers, and UK citizens. An undertaking of this complexity inherently carries limitations.

The project required an adaptive approach at each stage, with significant elements of learning by doing. Many of the methods, processes and analytical frameworks were developed and refined during delivery, reflecting both the innovative nature of the work and the interdisciplinary composition of the team. Adapting the deliberative process to different topics and modelling tools introduced variations across DMBs in both delivery design and data collection approaches.

Despite the two-year timeframe, time limitations still restricted capacity to address all relevant questions in sufficient depth. Tight timescales within DMBs imposed short analysis timelines, where longer timelines could enable deeper transcript analysis, richer narrative framing of citizen-led model inputs, and more time to produce complex sectoral and whole system model runs.

Limited time and availability of appropriate tools constrained capacity for panel members to consider trade-offs in real time and to explore wider and more dynamic scenario variation. This was mitigated by our development of a bespoke dashboard to enable members to interact with and explore a variety of elements of six whole-system scenarios.

Finally, limitations and their mitigations noted in both Barrett et al. (2022) [5] and Sharmina et al. (2025) [11] remain applicable.

## Methods-only bibliography

# Supplementary Information

# Deliberative panel process

## 7 Ethics, member safeguarding, and evaluation

Ethics approval was secured from the Lancaster University Faculty of Science and Technology and included securing informed consent from participants for the anonymised use of data provided and the secure and confidential handling of personal data.

The process was designed and delivered in collaboration with public participation charity Involve (www.involve.org.uk). As well as working with the research team to design the overall process and each individual workshop, and providing lead facilitation throughout the process, Involve also delivered participant support, safeguarding and welfare measures. Examples include providing a counsellor for members to access during and after discussions of potentially distressing themes, and providing IT support and equipment for participants that required them. Involve held all panel member personal data, including contact details, and were responsible for all contact with members for the duration of the panel.

The process was evaluated by an independent evaluator. Feedback was sought from panel members after each DMB, and periodically from the panel's advisory group, speakers, and modelling team. The evaluator had access to all provisional speaker lists, presentations, facilitation plans, recordings, and advisory group meetings. She attended panel meetings in DMBs 2, 6 and 8. Advice and feedback was provided on an ongoing basis by email, online and in-person meetings, and interim reports following each DMB. The final report is available at https://a-b-consulting.co.uk/2025/11/03/case-study-positive-low-energy-futures/.

## 8 Panel member selection

Panel members were initially recruited through sortition to ensure representativeness, by the independent Sortition Foundation (https://www.sortitionfoundation.org/). Selection factors included age, gender, occupation type, ethnicity, disability, rural or urban residence, concern about climate change, and political views. These factors all have a direct or indirect bearing on how EDR policies and practices would affect people, and/or how they are likely to respond to EDR proposals. Political views have not often been a selection criterion for citizens' assemblies, leading to political imbalance, and are increasingly important in shaping views on climate-related policies given the fragmentation of political consensus [1]. Additionally, all members lived within 90 minutes travel time of central Manchester (by car or public transport), to facilitate in-person meetings. The long-term nature of the panel inevitably led to attrition of panel members due to changing life circumstances. As a result, seven (7) new panel members were recruited between DMBs 4 and 5, using focused market research selection methods to fill the specific criteria-based gaps left by departing members. New members attended a one-day in-person event at which project team members briefed them on the panel process, provided an abbreviated introduction, explained the topics discussed so far and their key outcomes, and answered questions. Five existing panel members also attended this event to provide members' perspectives and to enhance the process of welcoming and helping new members feel part of a project in which other participants were already established.

## 9 Deliberative Modelling Blocks - overview

Table SI1 sets out the timeline for the DMBs; the theme, "calling question" (the central, overarching prompt that frames the core problem and guides participants in their deliberations), and specific topics discussed in each DMB; and the model used to analyse the outputs, where applicable.

The structure and flow of DMBs is shown in Figure SI1.

Table SI1: Overview of Deliberative-Modelling Block content, timeline and associated models

| DMB | Date | Theme | Calling question | Topics discussed | Model used to analyse data |
|---|---|---|---|---|---|
| 1 | Apr 2024 | Introduction | As a nation we need to use less energy, to meet our climate targets, increase energy security, and save households money. How can we do this in ways that work for everyone? | • Climate science and impacts<br>• Carbon targets and budgets<br>• Modelling and scenarios, including previous EDR work<br>• Key sectors<br>• Panel process and focus | Data arising from DMB 1 did not contribute to modelling, but helped guide the orientation of later DMBs |
| 2 | Jun-Sep 2024 | What we eat | What could a healthier and more sustainable national diet look like, what may prevent or support us reaching that, and how we can ensure that changes work for everyone? | • Reducing meat and dairy consumption<br>• Reducing food waste<br>• Increasing local food production | Hybrid physical input-output model |
| 3 | Sep-Nov 2024 | What we buy | How could we buy things in ways that use less energy, whilst supporting everyone to live a good quality of life? | • Car clubs and journey sharing<br>• Libraries of things<br>• Longer-life, repairable electronics<br>• Conscious clothing consumption | Hybrid model (UK MRIO + bespoke construction sector model) |
| 4 | Jan-Mar 2025 | How we heat our homes | How could we reduce energy demand from heating the nation's 1homes in ways that are fair, save people money, and improve comfort and health? | • Heat pumps<br>• Insulation retrofit<br>• Flexible energy use tariffs | Cambridge Housing Model |
| 5 | Jun-Jul 2025 | How we get around | How could we reduce energy demand from moving around, while ensuring that everyone is able to meet their needs for travel? | • Car size<br>• Modal shift<br>• Reducing trip frequency<br>• Reducing trip distance | Transport Energy Air pollution Model for the UK (TEAM-UK) |
| 6 | Sep 2026 | Whole system implications | What does a reasonable and fair balance between the three pillars (energy demand reduction, clean electricity generation and | • Clean power generation<br>• Land-based and engineered carbon dioxide removal | UK TIMES |
| | | and trade-offs | carbon dioxide removals) of our energy future look like for a net zero UK? | • The balance between the ‘3 pillars’ of net zero. | |
| 7 | Nov 2026 | Panel-selected topics: Place-based visions; aviation | Flying is highly valued, but very difficult to decarbonise. Should we manage demand for flying, by how much, how, and for who?<br><br>What does a positive low energy future look like for different kinds of places in the UK? | • Limiting aviation growth<br><br>• How different types of settlement might reduce energy demand in response to scale-specific pressures and drivers | Transport Energy Air pollution Model for the UK (TEAM-UK)<br><br>Place-based visions contributed to a more detailed understanding of conditionality, but were not modelled |
| 8 | Feb 2026 | Final Citizens’ Panel Scenario review and principles development | As a nation we need to use less energy, to meet our climate targets, increase energy security, and save households money. How can we do this in ways that work for everyone? | • Challenges of significantly increasing land-based CDR<br>• Accuracy of scenario in reflecting panel discussions and expectations<br>• Alignment of scenario with member views<br>• Principles for reducing energy demand fairly and effectively | UK TIMES |

Figure SI1: Structure and flow of Deliberative-Modelling Blocks

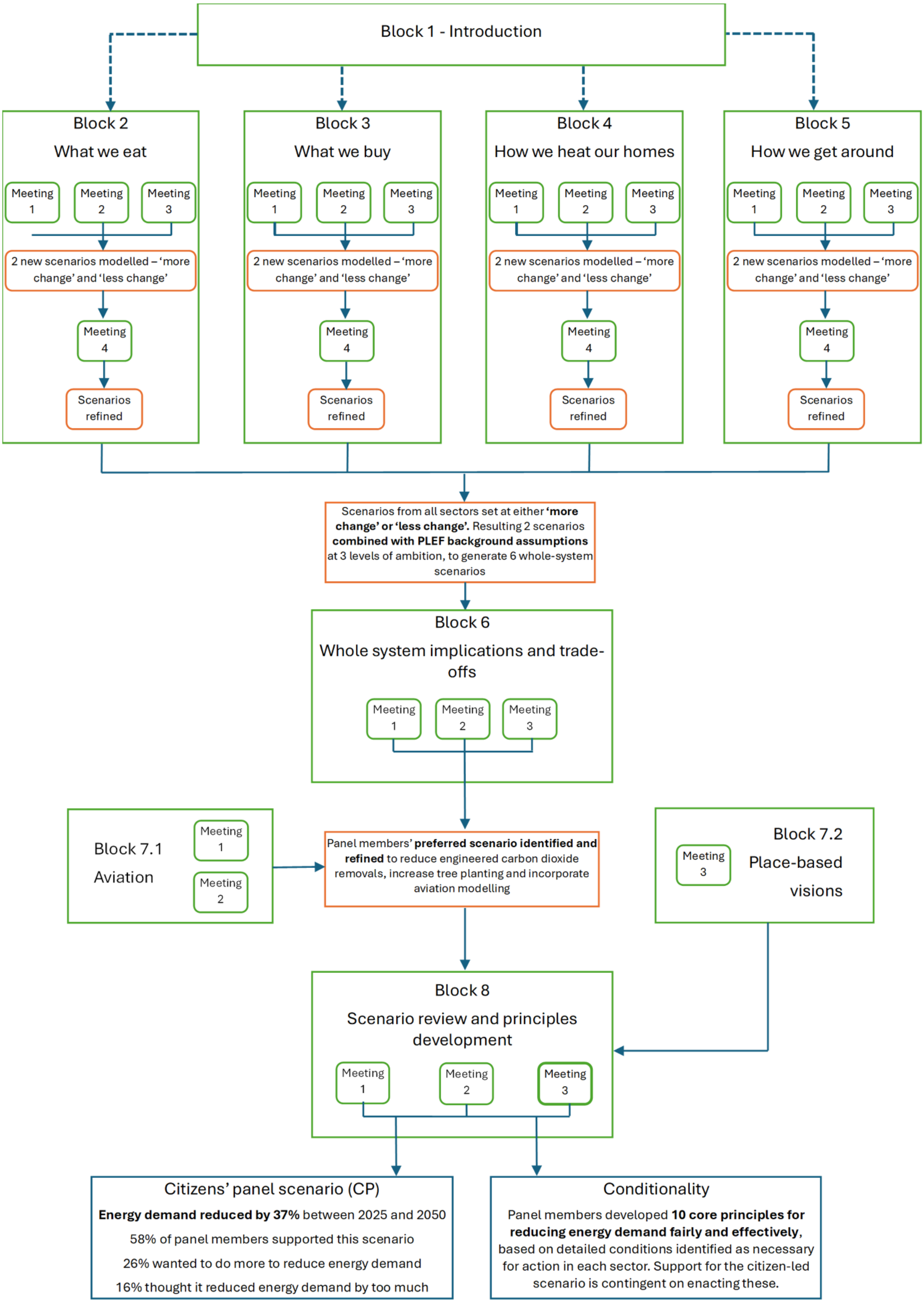

# 10 Deliberative Modelling Blocks – sample agendas and exercises

Below we provide examples of DMB meeting agendas and deliberation exercises (taken from DMB 6) to add indicative depth to the overviews presented above. We also provide an online facilitation plan template. All online and in-person meetings were planned and delivered using this template, with minor variations to reflect operational differences for in-person meetings. Online meetings were conducted using Zoom. Copies were shared in advance with all table facilitators. Facilitators were additionally briefed in advance of each meeting and debriefed afterwards to enable continuous improvement and for the coordination team to address any issues arising as soon as possible. Ongoing input from the independent evaluator also assisted with continuous improvement and addressing issues arising. Detailed facilitation plans remain the property of The Involve Foundation.

## 10.1 Example meeting agendas: DMB 6

### *10.1.1 Meeting 1 (2 hours online)*

**Aims and objectives for meeting 1**

- To provide information on block scope: introduce and explain the calling question, discussion topics, block agendas and discussion sequencing
- To introduce the dashboard, whole-system modelling, and the three pillars of pathways to net zero
- To let members ask questions to support understanding of the basics on the above

**Outputs from this meeting**

- Member hopes and uncertainties for this block
- Member questions on the dashboard, modelling, and three pillars

**Agenda: Section title and purpose**

1. **Introduction**
   Welcome people to the process, remind them of purpose of panel, let them know what to expect in this meeting

2. **Information provision - scope of the block**
   Clarifying scope of the block and the idea of the interplay between the three pillars

3. **Warm Up**
   Members reconnect and reflect on differences between this and previous blocks, and any uncertainties and hopes they have

4. **Information provision - what the whole system model shows**
   Members see how the panel-led sectoral modelling has been translated into a model representing the whole energy system, and what that model can do

5. **Rapid question generation**
   Members consider the information and generate questions

6. **Break**

7. **Information provision - Introducing the three pillars**
   Members are introduced to EDR, CDR, and CES as three core interacting 'pillars' of the route to net zero, emphasising trade-offs between them, ie doing less of one requires doing more of others

8. **Information provision - Introducing the dashboard**
   Members see the dashboard for the first time. Walk through demonstration of what it shows and how to compare between scenarios, with worked examples. Focus on how to interpret charts and what kind of information is important (eg trends not details, comparisons between scenarios rather than precision about individual scenarios)

9. **Reflections**
   Generating questions for all speakers

10. **Plenary Q&A**
    Members get the chance to ask outstanding/interesting questions to support understanding

11. **Wrap up**
    Members leave feeling valued and knowing what will happen next

### *10.1.2 Meeting 2 (2 hours online)*

**Aims and objectives for meeting 2**

- To provide information on CDR as one of the pillars of net zero
- To give members time to apply information on CDR to their values, opinions, and preferences around routes to net zero
- To have plenty of time to ask questions to support member's understanding on carbon dioxide removals.

**Outputs from this meeting**

- Member questions on CDR
- Member reflections on what is important to them about CDR

**Agenda: Section title and purpose**

1. **Welcome**
   Welcome members back, explain what is going to happen in today's meeting

2. **CDR information provision**
   Members understand what CDR is, basics of risks and opportunities, and what the dashboard shows on CDR

3. **Reflections and question generation**
   Members reflect on what they heard and generate questions for the Q&A

4. **Plenary Q&A part 1**
   For members to have their questions answered by speaker and additional expert panellists

5. **Break**

6. **Follow up and clarification question generation**
   A chance for members to think about follow-up questions, addressing gaps in understanding or areas to probe further

7. **Plenary Q&A part 2**
   For members to have their questions answered by speaker and additional expert panellists

8. **Reflections**
   For members to interrogate risks and opportunities in CDR and consider how this can differentially impact on groups of people or places. To begin to apply this to what it might mean for getting to net zero and the role of CDR alongside EDR

9. **Wrap up**
   Thanks, reminders for the next session

### *10.1.3 Meeting 3 (6.5 hours, in person)*

**Aims and objectives for meeting 3**

- To provide information on implementing a clean electricity system (CES) as one of the pillars of net zero
- To have plenty of time to ask questions to support member's understanding of a CES
- To give members time to apply information on CES to their values, opinions, and preferences around routes to net zero
- For members to use the dashboard to support weighing up where they feel the emphasis should be between the three pillars
- To deliberate on the balance of the three pillars and the values and priorities that lie behind that, including what would be considered acceptable trade-offs
- To vote on the balance of emphasis between CDR, EDR and CES
- To identify and prioritise block 7 topics

**Outputs from this meeting**

- Member questions on CES
- Member reflections on what is important to them about CES
- Member reflections on the balance of the three pillars, having used the dashboard to engage with implications
- Members views on key considerations for a pathway to net zero (cost; risks and challenges to delivery; impacts on everyday life; land use; energy security; timescales; economy; environmental and health impacts) as they apply across the three pillars
- What members most want to see from pathways to net zero and what members most want to avoid from pathways to net zero
- Where members would put more or less emphasis on EDR, CDR, and CES

**Agenda: Section title and purpose**

1. **Introduction**
   Make people feel welcome, ensure everyone knows what's happening, and who is who

2. **Warm up and check in and active listening**
   Members reflect on the block so far and do so by practicing looping for understanding

3. **CES information provision**

Members understand what the electricity system is, where electricity comes from now and where it may come from in a future CES, the scale of electricity generation, transmission and distribution needed, and the features and challenges of different types of CES generation

4. **Reflections and question generation**
   Members reflect on what they heard and generate questions for the Q&A

5. **Plenary Q&A part 1**
   For members to have their questions answered by speaker and additional expert panellists

6. **Break**

7. **Follow up and clarification question generation**
   A chance for members to think about follow-up questions, addressing gaps in understanding or areas to probe further

8. **Plenary Q&A part 2**
   For members to have their questions answered by speaker and additional expert panellists

9. **Reflections on CES**
   For members to interrogate concerns, challenges and benefits of different types of CES infrastructure, including how this can differentially impact on groups of people or places. To begin to apply this to what it might mean for getting to net zero and the role of a CES alongside EDR and CDR

10. **Dashboard use**
    To demonstrate use of the online dashboard to explore the implications of different levels of EDR ambition for CES and CDR, emissions reduction, land use, and investment needed; and for members to explore the dashboard in groups of 3-4 via guided exercises and self-discovery. Facilitators will work with 2 groups each and members of the research team are available to answer questions.

11. **Lunch**

12. **Deliberation on the balance of pillars**
    For members to explore what they would want to see, and would want to avoid, from a pathway to net zero in terms of a list of key considerations (cost; risks and challenges to delivery; impacts on everyday life; land use; energy security;

timescales; economy; environmental and health impacts). For members to consider what this might mean for how much effort the nation dedicates to EDR, CDR, and CES. Headline information about the features of CDR, EDR, and CES in relation to each area is provided in member workbooks.

**13. Menti vote and reflection**
Vote on degree of emphasis to be placed on each of EDR, CDR, and CES. Reflections on specific aspects any of the pillars and nuances that may be lost or concealed by voting

**14. Break**

**15. Block 7 topic choices**
To identify and prioritise topics for discussion in Block 7, following information provision on pre-prepared options based on discussions in this and previous DMBs, and/or usefulness for analysis (with additional option for 'anything else')

**16. Voting, feedback and wrap up**
Briefing on next steps and time for members to complete evaluation survey, including additional voting on direct trade-offs between the 3 pillars (to triangulate with Menti vote), and on topics for next block

## 10.2 Example deliberative exercise

The following is an extract from the DMB 6, meeting 3 facilitation plan, sections 12 (deliberation on the balance of the pillars) and 13 (Menti voting and reflection).

**Section 12: Deliberation on the balance of pillars**

**Lead facilitator** Welcomes everyone back and explains that we are going to go into our full table groups again up to 3.05pm to spend time exploring:

> *We will need to have clean electricity, carbon dioxide removals and energy demand reduction to reach net zero. They can be combined in different ways but if we do less of one we need to do more of others. If it was up to you, based on the information you've been given across this block, where would you put more and less emphasis in policies to reduce emissions?*

Note that everybody who was available to answer questions in the dashboard session is still here and can come over to table groups to answer any questions you have.

**Table facilitators**

**Start recording**

1. Looking at the things to consider in your workbook, let's discuss what this might mean for this balance of more or less emphasis between energy demand reduction, the clean electricity system, and carbon dioxide removals.

   **Facilitator note**: *workbooks contain a table summarising headline information provided on each pillar with regard to each of the areas of consideration, i.e. cost; risks and challenges to delivery; impacts on everyday life; land use; energy security; timescales; economy; environmental and health impacts.*

   Let members select a consideration area to begin with.

2. Write the selected consideration area at the top of a blank flipchart page - on the left draw a continuum from 'What we would want to see' to 'What we want to avoid' and invite members to share what they would and would not want to see in getting to net zero in relation to this area. Work with members to explore these considerations in more detail and going beyond the table headings to more granular discussion of what they would want/not want to see or accept as a consequence and why.

Use post it notes to add considerations to the flipchart. Draw out any points of interest, for example ask members whether similar considerations should be grouped together or where something similar seems to sit at different points on the continuum etc.

As members add post it notes, support them to explore why they are placing them in that place on the continuum, why that matters and whether the group agrees (no need to overly worry about precise placing, more important to have the conversation on acceptability more generally).

**Facilitator note** - *In some instances you will be able to support members to explore the table of considerations in more detail. E.g. for the electricity system consideration 'Major impacts on countryside, nature and local communities (e.g. pylons, windfarms, storage)' you can get members to explore what they have seen and what some of the limits might be e.g. they may accept solar farms in some locations not others.*
*However where members are less able to consider what may sit behind a consideration in more detail, call over research team members to assist).*

3. After around 10 minutes, ask members what could make more acceptable the things they don't want to see.

**Facilitator note** - *for example, if they dislike higher taxes, could it be that higher taxes are more acceptable if they were linked to value of assets or pollution etc.* Depending

| |
|---|
| *on the flow in your group you may integrate this more closely with when members place considerations on the continuum.*<br><br>Aim to spend ~ 20 mins per consideration area and explore at least 3 consideration areas repeating steps 2 and 3 above. As you complete your flip charts, pin to the display board. Call over **support** to help as needed with arranging flips!<br><br>**Facilitator note** - *Some exploratory questions you may want to draw on to support members in this session to orient towards important aspects:*<br><br>• *Would some people be more affected by that than others? How?*<br>• *Do you think any positive impacts could be spread or increased? How?*<br>• *Do you think any negative impacts could be reduced or avoided? How?*<br><br>**Stop recording**<br><br>**Lead facilitator** reminder for all when 15 minutes left<br><br>Calls everyone back to plenary. |
| **17. Section 13: Menti vote and reflection**<br><br>**Lead facilitator** shares Menti slide and explains that the Menti asks you to distribute 100 points between the three pillars - the higher numbers you give to a pillar the greater emphasis you would like to see on that pillar.<br><br>You can distribute your 100 between the three pillars however you like but you have to use all 100 - this is because we have to get to net zero.<br><br>Invites everyone to vote.<br><br>**Table facilitators** support voting if needed. Lead facilitator and support team can also be called over if needed.<br><br>**Lead facilitator** shows the results and hands back to table facilitators for a quick reflection before the break.<br><br>**Table facilitators** - support your members to draw out any reflections on the voting process or results.<br><br>Any nuance that got lost - eg. were there parts of some pillars you really like/dislike that made this voting hard?<br><br>Capture on flip (direct writing or post its as suits)<br><br>NB - members will also be able to explain this in the feedback form - inc what parts of the pillars you like/dislike. |

**Lead facilitator** sends to break - Reminder that Anna [external evaluator] will want to chat with some of you informally - no obligation to do so if you need to spend the time in the breaks in a different way to suit you.

## 10.3 Template facilitation plan

# Our Energy Futures, Block x, session x. Facilitation plan.

**Aims & objectives:**

- Xxxxxxx
-

**Meeting outputs:**

- xxxxxxx
-

**Zoom Joining Details**

Join Zoom Meeting

xxxxxxx

Meeting ID:

Passcode:

Link to register: xxxxxxx

**Groups & links to online whiteboards and notes facilities:**

| **Facilitator** | **Members** | **Jamboards** | **Google notes docs** |
|---|---|---|---|
| xxxxxxxx | xxxxxxxx | xxxxxxxx | xxxxxxxx |
| xxxxxxxx | xxxxxxxx | xxxxxxxx | xxxxxxxx |
| xxxxxxxx | xxxxxxxx | xxxxxxxx | xxxxxxxx |
| xxxxxxxx | xxxxxxxx | xxxxxxxx | xxxxxxxx |
| xxxxxxxx | xxxxxxxx | xxxxxxxx | xxxxxxxx |

**Recording, transcription and storage protocols:**

Xxxxxxxxxxxxxxx

| Start - Finish time & duration | Session and Purpose | Facilitation notes | Resources required | Outputs | Production notes |
|---|---|---|---|---|---|
| | | | | | |
| | | | | | |
| | | | | | |
| | | | | | |
| | | | | | |
| | | | | | |
| | | | | | |
| | | | | | |
| | | | | | |
| | | | | | |

Post-meeting tasks

- Xxxxxxxx
-

## 11 Variations to the deliberative process

To introduce a further element of co-creation and panel agency, DMB 7 themes were selected by panel members themselves at the end of DMB 6. Members were presented with a shortlist of topics developed from a combination of researcher interest, panel member interest gauged from previous discussions, and relevant knowledge gaps, with a brief introduction to each topic and why it was considered to be of interest. Members also had the opportunity to propose entirely new topics. Facilitated, small group discussion sessions enabled members to discuss and prioritise areas of interest, which were then subject to individual voting by the whole panel as part of the regular online feedback survey.

The discussions on aviation informed further modelling, and contributed to the modelled Citizens' Panel scenario. The place-based visioning exercise did not contribute to the modelling, but helped with understanding of how conditionality could play out in different types of settlement.

In DMBs 7 and 8, panel members had access to a 'drafting table' - members of the research team who they could call on to review and assist with the development of place-based visions and EDR principles. The role of the drafting table was to respond to questions of clarification, ask probing questions, for example about implications of proposals or trade-offs that panel members may want to consider, and to provide expert knowledge to ensure the practicality of proposals, but not in any other way to guide or influence panel members' proposals.

## 12 SI Bibliography